\documentclass[sigconf]{acmart}
 
\usepackage{url}
\usepackage{enumitem}
\usepackage{mwe}
\usepackage{multirow}
\usepackage{multicol}
\usepackage{array}
\newcolumntype{P}[1]{>{\centering\arraybackslash}p{#1}}
\newcolumntype{M}[1]{>{\centering\arraybackslash}m{#1}}
\usepackage{changepage}
\usepackage{subcaption}
\usepackage{cleveref}

\emergencystretch=\maxdimen
\AtBeginDocument{%
  \providecommand\BibTeX{{%
    \normalfont B\kern-0.5em{\scshape i\kern-0.25em b}\kern-0.8em\TeX}}}

\copyrightyear{2022} 
\acmYear{2022} 
\setcopyright{acmcopyright}
\acmConference[ICSE-SEET '22]{44nd International Conference on Software Engineering: Software Engineering Education and Training}{May 21--29, 2022}{Pittsburgh, PA, USA}
\acmBooktitle{44nd International Conference on Software Engineering: Software Engineering Education and Training (ICSE-SEET '22), May 21--29, 2022, Pittsburgh, PA, USA}
\acmPrice{15.00}
\acmDOI{10.1145/3510456.3514142}
\acmISBN{978-1-4503-9225-9/22/05}

\begin{document}

\title[Interactive Web-Based Authoring and Playback Integrated Environment] {ITSS: Interactive Web-Based Authoring and Playback Integrated Environment for Programming Tutorials}

\author{Eng Lieh Ouh}
\email{elouh@smu.edu.sg}
\affiliation{%
  \institution{Singapore Management University}
  \country{Singapore}
}

\author{Benjamin Kok Siew Gan}
\email{benjamingan@smu.edu.sg}
\affiliation{%
  \institution{Singapore Management University}
  \country{Singapore}
}

\author{David Lo}
\email{davidlo@smu.edu.sg}
\affiliation{%
  \institution{Singapore Management University}
  \country{Singapore}
}

\renewcommand{\shortauthors}{Author1 and Author2 and Author3}

\begin{abstract}
Video-based programming tutorials are a popular form of tutorial used by authors to guide learners to code. Still, the interactivity of these videos is limited primarily to control video flow. There are existing works with increased interactivity that are shown to improve the learning experience. Still, these solutions require setting up a custom recording environment and are not well-integrated with the playback environment. This paper describes our integrated ITSS environment and evaluates the ease of authoring and playback of our interactive programming tutorials. Our environment is designed to run within the browser sandbox and is less intrusive to record interactivity actions. We develop a recording approach that tracks the author's interactivity actions (e.g., typing code, highlighting words, scrolling panels) on the browser and stored in text and audio formats. We replay these actions using the recorded artefacts for learners to have a more interactive, integrated and realistic playback of the author's actions instead of watching video frames. Our design goals are 1) efficient recording and playback, 2) extensible interactivity features to help students learn better, and 3) a scalable web-based environment. Our first user study of 20 participants who carry out the author tasks agree that it is efficient and easy to author interactive videos in our environment with no additional software needed. Our second user study of 84 students using the environment agrees that the increased interactivity can help them learn better over a video-based tutorial. Our performance test shows that the environment can scale to support up to 500 concurrent users. We hope our open-source environment enable more educators to create interactive programming tutorials.
\end{abstract}




\maketitle
\section{Introduction}
An interactive video that combines audio and visual elements is a video built to enable engagement beyond viewing. Zhang et al.  \cite{zhang2006instructional} study shows that interactive video enables engagement, participation, responsiveness, and active engagement of students. It allows students to pay full attention to educational material through active interaction. Another study by Preradovic et al. \cite{preradovic2020investigating} supports earlier findings that attention cues foster learning and cues in interactive videos such as the author's highlighting of text during explanation had a positive effect on the learning process and learning outcomes. 

Video-based tutorials have been shown to help novice programmers to learn to code. A study by Kim and Ko \cite{kim2017pedagogical} on pedagogical analysis of online tutorials highlights that active engagement of learners to interactively write code in tutorials is effective to learn to code. Our work is inspired by a paper from Bao et al. \cite{bao2018vt}. They study that by abstracting the author's interactive actions from the programming video tutorials, the learners can freely explore the captured workflows and interact with files, code and program output in the tutorial playback. Their user study shows that the system can help developers find relevant information in video tutorials better and faster, leading to a more satisfactory learning experience. Their approach utilizes operating-system (OS) level accessibility APIs to track the author's movements and capture contextual events so that users playing back the tutorial can have a better learning experience. Similar approaches of using OS accessibility APIs to record interactivity is also used in \cite{fraser2019replay,zhang2009smarttutor}. This paper extends the study to record the author's interactive actions in video tutorials. Instead of installing additional software to use the OS accessibility APIs, which can be intrusive, our approach uses browser in-built accessibility APIs and runs within the browser sandbox. The recording environment is also fully integrated with the playback environment, allowing a faster turnaround time to preview and test tutorials. 

The improvements of browsers accessibility capabilities over the years have increased the number of students learning using web-based services \cite{berns2019myr, valez2020student} and they find it more convenient. Our approach is to design using these capabilities on a web-based integrated environment for authoring and playback. During authoring, We record interactivity actions by the author, such as scrolling and re-positioning panels, highlighting text/code on any panels, and typing and executing code. The author can highlight words and code concurrently to better relate the flow between the contents and coding. We replay these actions as if the author is performing these actions concurrently along a timeline.  We hope to provide an improved and realistic learning experience for the students. 

The application of interactive video within the educational process is still not sufficiently explored \cite{preradovic2020investigating} and one reason cited is educators lack of access to tools to implement interactivity within a video. Our interactive web-based integrated environment is open-source (openly available via our public GitHub repository \cite{ITSSGitHub} with Apache 2.0 open source license) and can fill this gap.

This paper describes ITSS an interactive web-based integrated environment for authors to create interactive programming tutorials for students to playback. 
In designing ITSS, we have the following main goals:
\begin{enumerate}
\item The environment should allow the authors to \textbf{efficiently} record the interactive tutorial on the browser.
\item The environment should be \textbf{extensible} to allow tutorials of new interactive features and programming languages to help students learn better.
\item The environment should be \textbf{scalable} to ensure the learning experience is not affected by the growth of user numbers.
\end{enumerate}

In this paper, we describe our design and implementation in Section 2. The authoring and playback flow is described in Section 3, and Section 4 describes other related work. We analyze our user studies and performance tests in Section 5 and conclude in Section 6. This paper makes the following contributions:
\begin{enumerate}
\item We describe the design of the ITSS server, an integrated, extensible and scalable environment for authoring and playback of programming tutorials.
\item We present our novel recording approach to capture and playback the author's interactive actions.
\item We present findings of two user studies to evaluate our environment for authoring and playback of programming tutorials and performance tests to scale our environment.
\end{enumerate}

\section{DESIGN AND IMPLEMENTATION}
ITSS environment comprises the following four key components:
\begin{enumerate}
\item The \textbf{ITSS web client} is responsible for rendering the web interface components, built based on React framework.
\item The \textbf{ITSS services server} is responsible for processing the business logic and designed based on microservices style. There are currently over 20 microservices implemented.
\item The \textbf{ITSS database repository} is responsible for persisting the tutorial recordings metadata and environment data.
\item The \textbf{ITSS recording file repository} is responsible for persisting the recording artefacts created for each tutorial.
\end{enumerate}

\begin{figure*}[ht] 
\centering
\includegraphics[width=380pt]{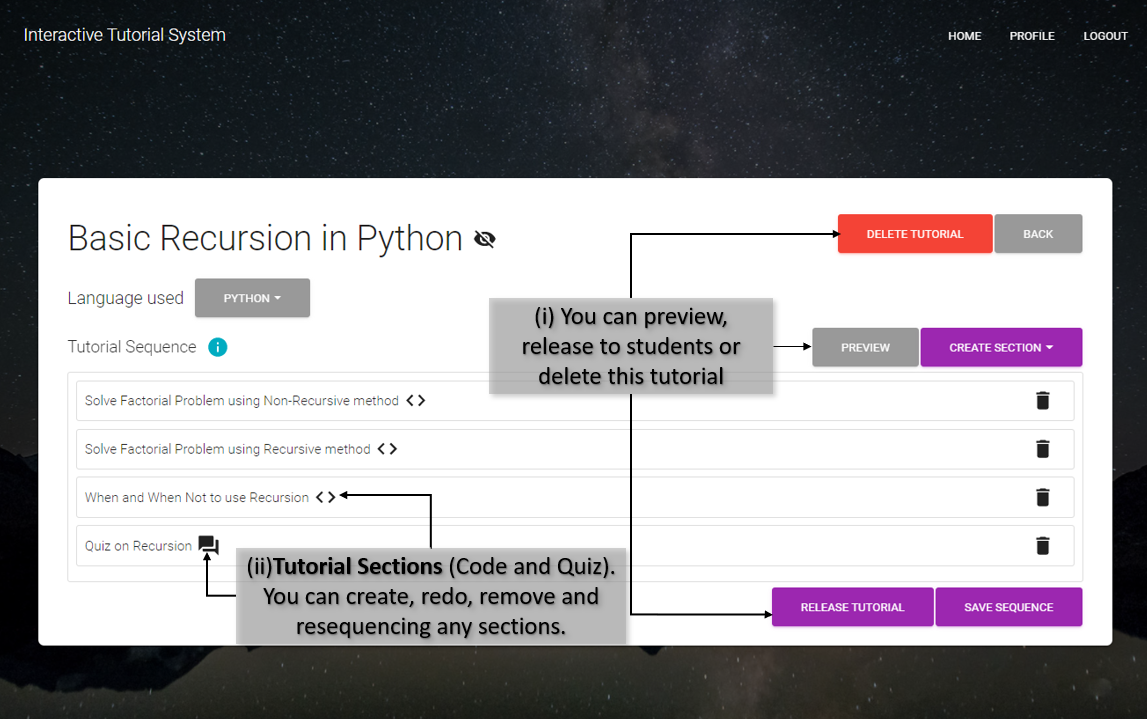}%
\caption{Tutorial Sections Screen}%
\label{subfiga}%
\end{figure*}

One design goal is that authors can \textbf{efficiently} record interactive tutorials. Our environment allows authors to record quickly on the browser for playback without additional software setup to deliver the learning outcomes of the tutorials. The author can record two types of tutorial sections – coding and quiz, as shown in Figure \ref{subfiga}. For the rest of this paper, we focus on the coding section which is panel-based and contains our interactivity features. Our coding section design allows for bite-sized tutorial sections. The author can record each tutorial section separately for each learning outcome, add, modify, delete, interleave and re-sequence them . 

Another design goal is the \textbf{extensibility} of the environment. The panel-based design allows additional interactive features to be integrated easily. We currently integrate Markdown, Ace IDE \cite{AceEditor} and PythonTutor \cite{guo2013online} in separate panels. Markdown language allows text-based editing and conversion on the fly. We reuse the rich Ace IDE coding features and extend with input/output panels and code execution features. We support Python and Java and are extendable with a language-specific back-end service. The extensions required to support a new compile language are installing additional language-specific libraries and a script to compile and execute the code. If we want to extend with an interpreted language, we can modify the front-end and do not impact the back-end. We include PythonTutor to allow visualization of the code execution. We track the author's interactivity actions in text-based files, allowing us to modify/extend the tutorial contents without re-recording and possibly other interactions in the future. Our environment incorporate the design principles of Mayer \cite{mayer2002multimedia, mayer2014multimedia} and  Griffin et al. \cite{griffin2009podcasting} for students to have good retention and transfer performance. The playback of the audio, notes or other panels are synchronized and panels design allow more focused attention on a specific learning outcome.

Another design goal is the \textbf{scalability} of the environment. The public hosting of the ITSS environment is on Amazon Web Services (AWS) cloud. We use AWS Elastic Beanstalk to manage the deployment of the service, AWS Relational Database (RDS) for database repository and S3 for recording file repository. This architecture design allows us to utilize the auto-scaling feature of AWS to support increased workload. We externalized the environment configurations so we can port this environment to host locally, on other remote servers or other cloud platforms.

\begin{figure*}[ht] 
\centering
\includegraphics[width=380pt]{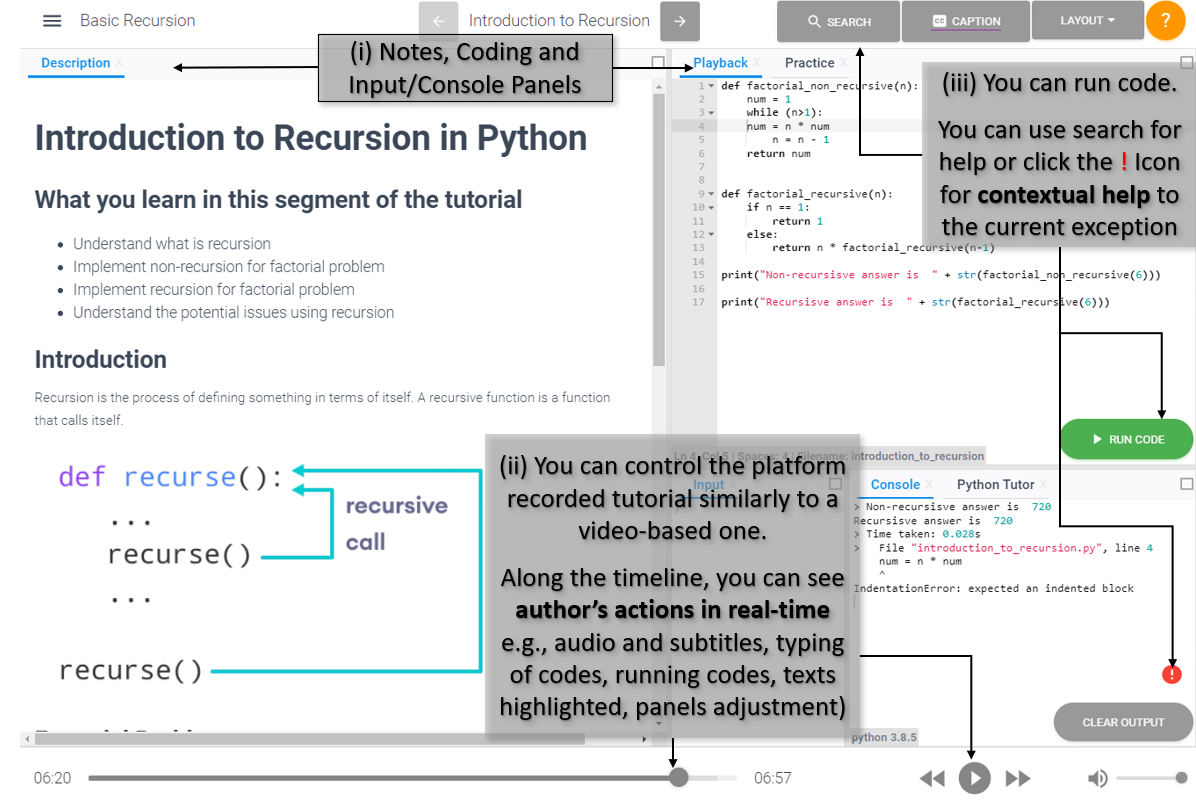}%
\caption{Tutorial Playback Screen}%
\label{subfigb}%
\end{figure*}

\vspace{-0.5em}
\subsection{Recording Approach}
Our recording approach differs from other video-based learning as we record in text and audio formats instead of the video frames. Specifically, we record the tutorial in the following artefacts. 
\begin{enumerate}
\item Author's audio. This artefact is currently stored in mp3 format and also transcribed into a text file for subtitles.
\item Author's notes. The notes are stored in Markdown format and converted to HTML on the browser.
\item Author's coding. The complete source code and the timing of the coding keystrokes are stored in JSON format.
\item Author's panel actions. The actions-related data such as texts highlighted, code typed, panel scrolling and positioning are stored with the timings of these actions in JSON format. 
\item Quiz section. The set of questions, answers, explanations and points are stored in JSON format.
\end{enumerate}
\vspace{-1em}
\subsection{Benefits and Trade-offs of our Design}
\label{sec2.2}
Our recording approach tracks interactive actions for playback to provide a realistic learning experience. There is no frame by frame to play each video but a more natural presentation of the author's interactions. Following are other benefits of our design:
\vspace{-0.7em}
\begin{enumerate}
\item The text-based artefacts such as the code, description can be modified separately without re-recording the entire tutorial.
\item Our environment can search the text-based artefacts accurately instead of reverse engineer from video frames.
\item Our environment can provide context-specific assistance to the student. For example, in Figure \ref{subfigb}, recommend StackOverflow postings that are context-specific to the Python "Indented block expected" exception message.
\item The data size of our recorded artefacts is smaller than a recorded video which reduces network latency and storage. The size of an uncompressed and compressed one minute video is about 67MB and 5MB. The total file size of a set of ITSS recorded artefacts is about 1.2MB. If we compress the mp3 file which accounts for over 95\% of the file size,  the size of the ITSS recorded artefacts is reduced to 0.4MB.
\end{enumerate}

There are design trade-offs to our recording approach. We are limited to recording browser events and contents from the same domain due to sandbox constraints. For example, we are unable to track events within the PythonTutor iframe as our environment invokes this service on another domain. We can overcome this limitation by implementing the PythonTutor framework within our environment, which we leave as part of future work.

\section{Authoring and Playback Flow}
An author can choose to register/login with email or use Google Login authentication, which allows us to use our university credentials via Google workspace accounts. Once logged in, the author can create a new tutorial of a specific programming language or edit an existing tutorial. Each tutorial is made of one or more sections, as shown in Figure \ref{subfiga} which the author can break down the content for bite-sized consumption. The author can create, redo, remove or re-sequence tutorial sections. The author can also preview, release the tutorial to students or delete the tutorial. If the author chooses a coding section, the panel-based screens are shown to present notes and write code. The authoring screen is similar to the tutorial playback screen as shown in Figure \ref{subfigb} with additional buttons to record. The author can upload existing content for each tutorial. The author can also create and preview new content online using the provided markdown editor. Once the author starts recording, our environment begins listening to the author's audio and continuously tracking the author's interactive actions along the recording timeline.  The author can type, compile and execute code with interactive inputs and outputs. If the author wants to emphasise key concepts, the author can scroll or maximise the panel and highlight the text content that is being explained. The browser will seek user permission to use the microphone if the author uses this browser to record for the first time. When the recording is completed, the author can preview, save or discard the these interactive actions of the recording. 

If a student login, the student is presented with a list of released tutorials on the homepage. Once a tutorial is selected, the student is presented with the tutorial playback screen as shown in Figure \ref{subfigb}. Part (i) of the figure shows the notes, coding, inputs and outputs panels the author initially configured. The student can remove, adjust the panel size, re-position the panel, save the layout to suit the learning needs and navigate between the tutorial sections. The student can play the tutorial, rewind, forward or control the author's replayed actions along the timeline as shown in Figure \ref{subfigb} part (ii). Each step of the actions is interactively executed on the student's browser, including the author's movement of panels, text highlight and undoing actions such as delete code. The student can concurrently practice typing and executing code as indicated in Figure \ref{subfigb} part (iii). The author may type out the entire code or provide the initial code to practice. In these cases, the student can copy the code at any point of the timeline, paste it onto the practice panel and execute it on the browser. The student can also search the tutorial for keywords or the environment can assist with encountered exceptions based on the text-based artefacts.
\section{RELATED WORK}
We categorized three areas of work of how instructors teach students to learn programming (1) Self-guided instructions, (2) Traditional videos and class quizzes (3) Online Coding tutorials. 

For (1), instructors prepare notes for students to self-learn. It can be in an interactive format such as Jupyter notebooks \cite{van2020jupyter}, allowing students to go through bite-sized learning notes, compile and see the intermediate results using a client-server architecture. For (2), instructors prepare traditional frame videos for students to watch and further practice in class. We acknowledge that faculty and students should explore and discover their preferences in authoring and learning. Self-guide instructions such as the Jupyter notebook allows students to learn at their own pace and can be integrated for authors in our environment. We cannot replace the benefits of educators' class interaction, but educators can use this environment to create more interactive tutorials. 

For (3), instructors select online coding tutorials for students to learn. A comprehensive list can be found in the work of Kim and Ko \cite{kim2017pedagogical}. The environments of these online coding tutorials contain features to play video tutorials and may involve interactive coding and executions. In comparison, ITSS focuses to provide more interactive features (e.g., words highlighting, panels scrolling) besides interactive coding. We design these features to integrate seamlessly and do not appear as silos. For example, authors can highlight contents and code, scroll, and adjust panel positions concurrently along the timeline. Two specific related works that also promotes interactive coding are Scrimba \cite{Scrimba,lauvaas2018teaching} and Codio \cite{croft2019computing}. To the best of our knowledge, Scrimba focuses primarily on front-end and not back-end technologies. Due to course needs, ITSS supports back-end Java and Python executions which require additional server components to enable that. Codio uses interactive playback to review student's submitted code. ITSS extends interactivity capability to the author's authoring and recording more interactive features beyond coding actions.
\section{Evaluations}
We conduct two user studies (authors and students separately) and a system performance test to evaluate the ease of authoring and playback, learning experience, usability and scalability of this environment. These factors affect the on-boarding of authors and impact the students' learning effectiveness. Our first study question (SQ1) seeks to understand from authors the ease to record using our environment and their opinions on the authoring features. Our second study question (SQ2) seeks to understand from students whether the increased level of interactivity helps them to learn programming better and their opinions on these interactivity features. Our third study question (SQ3) seeks to understand whether the system can handle up to 500 concurrent users. Institutional Review Board (IRB) approval is obtained from our university for the above study designs.
\vspace{-0.5em}
\subsection{USER STUDY 1 Design (Authors)}
We design this study to be anonymous and do not ask for participants' identifiable information. This study is unmoderated to minimize the influence of the research team. This study involves using the Loop11 tool \cite{Loop11} which provides tasks instructions to the participants on the browser. Loop11 task is either an action task to do or a survey task to answer questions. The participant can proceed to perform the task and indicate either task abandoned or task completed anytime. We design two author's action tasks (AT) to gather inputs for the study.
\begin{enumerate} [label=\emph{AT\arabic*}]
\item Requires the participant to author a new programming tutorial and experience the authoring interactivity features. 
\item Requires the participant to re-sequence tutorial sections of an existing programming tutorial.
\end{enumerate}

\noindent We design two survey tasks ST1 and ST2 for SQ1 with the below questions.
\vspace{-0.5em}

\begin{enumerate} [label=\emph{ST\arabic*}]
\item What do you think in terms of ease of use to record a programming tutorial on the ITSS integrated environment? Please indicate how strongly you agree or disagree with each authoring feature in terms of the usefulness to author.
\item Please indicate how strongly you agree or disagree with the following statements based on System Usability Scale\cite{lewis2009factor}. 
\end{enumerate}

We invited 30 intended first-time users of this environment comprising of faculty who need to use this environment to author programming tutorials for their programming courses. We also invited senior undergraduate students who need to use this environment to author their own tutorials for their presentations. They are required first to watch a short video on how a traditional video tutorial is recorded. As compared to traditional video recording, we seek to evaluate whether it's easy to record on ITSS and their opinions on the interactivity features implemented. There is no incentive given and it is entirely voluntary for them to participate.

\vspace{-0.5em}
\subsection{User Study 1 Results (Authors)}
20 (10 faculty and 10 students) out of the 30 invited participants completed the user study at their own pace with an average of 26 minutes spent. All of them are successful in their task executions. 

Regarding ST1 for SQ1, all 20 participants either agree (35\%) or strongly agree (65\%) that it is easy to record on the browser without other software installation and commented:
\begin{adjustwidth}{0.5em}{0pt}
\emph{"able to record fast and re-record"}\\
\emph{"remarkably quick to put together an interactive video compared to what we have to do offline  (i.e. screencast, add voice, edit, upload)"}
\end{adjustwidth}
\vspace{0.5em}

Table \ref{tab:study-results-2} shows participants’ opinions on the authoring features implemented. Besides being able to code and record audio, the features to create bite-sized sections that can be re-sequenced are also appreciated by at least 95\% of the participants. 
Regarding ST2 for SQ1, the overall SUS score is at the 74th percentile. Although this range and score percentile is considered good \cite{lewis2009factor}, there is room to improve based on comments. 
\begin{adjustwidth}{0.5em}{0pt}
\emph{“It is not clear the diffs between the 2 different type of section.”}\\
\emph{“The expand button is confusing (looks like a checkbox).”}\\
\emph{"Release Button should be another colour, i feel"}\\
\emph{“recognition rather than recall is better–need to use standard icons!”}. \\
\end{adjustwidth}
\vspace{0.5em}
\begin{table}
\caption{\label{tab:study-results-2}Participants’ Opinions on Authoring Features}
\vspace{-0.5em}
\begin{tabular}{ | p{5.3cm} | M{0.6cm} | M{0.6cm} | M{0.6cm} |}
\hline
\textbf{Authoring Features-Total 8} & \multicolumn{3}{c|}{\textbf{ST1 on Usefulness}}\\
\cline{2-4}
\small SD/D - Strongly Disagree/Disagree, N - Neither, SA/A - Strongly Agree/Agree & SD/D & N & SA/A\\
\hline
Record scrolling and highlighting of notes&	0\%&	10\%&	90\% \\
\hline
Write code and execute the program&	    0\%&	0\%&	100\% \\
\hline
Record audio for the tutorial&	            0\%&	0\%&	100\% \\
\hline
Preview the tutorial section&	            0\%&	10\%&   90\% \\
\hline
Re-record the tutorial section&	            0\%&	20\%&	80\% \\
\hline
Create bite-sized tutorial sections&	    0\%&	5\%&	95\% \\
\hline
Adjust sequencing of the tutorial sections&	0\%&	5\%&    95\% \\
\hline
Record your face for the tutorial&	        40\%&	30\%&	30\% \\
\hline
\end{tabular}
\end{table}
\vspace{-1em}
\subsection{USER STUDY 2 Design (Students)}
We design this study for students taking an undergraduate information systems and computing course that covers software design and development. The students are taught by the faculty online over Zoom using slides and traditional videos for the first half of the semester. In the second half, we asked the students to view ITSS recorded programming tutorials as part of their weekly lesson.
We design two feedback questions FQ1 and FQ2 below for the second study question (SQ2).
\begin{enumerate} [label=\emph{FQ\arabic*}]
\item Do you think the interactivity features of a traditional video or the interactivity features implemented on the ITSS environment help students learn to program better?  Please indicate how strongly you agree or disagree with each playback feature in terms of the usefulness to learn better. 
\item Please indicate how strongly you agree or disagree with the following statements based on System Usability Scale \cite{lewis2009factor}. 
\end{enumerate}
\vspace{-0.5em}
We invited 88 first time users of this environment comprising undergraduate students taking the course. There is no incentive given and it is entirely voluntary for them to participate. There are no participants involved in both User Study 1 and 2.
\vspace{-0.5em}
\subsection{User Study 2 Results (Students)}
84 out of the 88 invited participants completed the feedback questions anonymously at the end of the semester. Regarding FQ1 for SQ2, 72\% of the students either strongly agree or agree that the interactivity features can help learn to program better and like this self-paced option. 20\% of them has no preference, while 8\% prefer traditional video. Many students with no preference or prefer traditional video miss the advanced video features such as speed controls which are currently lacking in our environment. Table \ref{tab:study-results-1} shows participants’ opinions on the interactivity playback features. Many commented on the coding interactivity features:
\begin{adjustwidth}{0.5em}{0pt}
\emph{"Having an interactive platform where I can both watch an instructional video and try out the code myself would help me to better remember concepts." } \\
\emph{"Being able to run it directly, make small changes, and introduce new test cases helps a lot with understanding." } \\
\emph{"I really enjoy how we are able to observe the author edit their code in real-time. This feature not only makes reading the code clearer but also allows us to copy/modify the code to our own liking."} \vspace{0.5em}

\end{adjustwidth}
The results for adding the author's face in the recording are split among the students, which is consistent with user study 1. Mayer \cite{mayer2014multimedia} and Pi et al. \cite{pi2017effects} argued that the inclusion of an instructor's image would not have a strong impact on learning. On the other hand, Kokoc et al. \cite{kokocc2020effects} found that picture-in-picture video lectures with the instructor's image yield improved performance. To minimize distraction, our environment continues not to show the instructor's image. It can be implemented when there is more evidence supporting it. A student participant summarised:
\begin{adjustwidth}{0.5em}{0pt}
\emph{Don’t think having the lecture face will affect my experience in using the system though”}.
\end{adjustwidth}
\vspace{0.5em}

The "Automated contextual help" and "Configure preferred layout of panels" features received less agreements than we expected. Based on comments, this outcome may be driven more by the interface design of this feature and not the benefits of these features. Some participants feel there are too many panels which can be overwhelming. For contextual help search, we display a red exclamation icon when an exception occurs as shown in Figure \ref{subfigb} and many participants are confused by this design and can be summarised by this comment
\begin{adjustwidth}{0.5em}{0pt}
\emph{"The exception red button is small and takes a while to figure out what is the use of it."}
\end{adjustwidth}
\vspace{0.5em}
Regarding FQ2 for SQ2, the overall SUS score is at the 75th percentile, similar to User Study 1. We acknowledge there is room to improve based on below comments. We seek to improve the interaction designs of this environment to better match users' mental models as suggested below for future work. 
\begin{adjustwidth}{0.5em}{0pt}
\emph{"dislike some component cannot use enter to confirm like search."}\\
\emph{“Can consider tracking mouse movements of instructor too.”}
\end{adjustwidth}
\vspace{0.5em}

During this user study, we realise there are errors in the recorded tutorials (e.g., missing code). We managed to correct them easily as the tutorial recordings are stored in text-based artefacts. We also encounter server resources being hogged due to infinite loops code. We mitigate by limiting the code execution to 10 seconds. There are also observations that although students prefer learning using ITSS, they express that recorded ITSS tutorials should be interleaved with synchronous lessons to interact with the faculty and not a replacement for the entire lesson. Alternatively, recorded ITSS tutorials can be used as pre-class / supplementary materials.
\vspace{-0.5em}
\begin{table}
\caption{\label{tab:study-results-1}Participants’ Opinions on Playback Features}
\vspace{-0.5em}
\begin{tabular}{ | p{5.3cm} | M{0.6cm} | M{0.6cm} | M{0.6cm} |}
\hline
\textbf{Interactive Playback Features-Total 9} & \multicolumn{3}{c|}{\textbf{FQ1 on Usefulness.}}\\
\cline{2-4}
\small SD/D - Strongly Disagree/Disagree, N - Neither, SA/A - Strongly Agree/Agree & 
\textbf{SD/D} & \textbf{N} & \textbf{SA/A}\\
\hline
Configure preferred layout of panels	& 10\%	& 23\%	& 67\% \\ 
\hline
Observe author's writing of the code	& 1\%	& 3\%	& 94\% \\
\hline
Observe author's highlighting of words	& 1\%	& 1\%	& 98\% \\
\hline
Copy author's code to practice	        & 1\%	& 2\%	& 97\% \\
\hline
Practice and execute your own code	    & 0\%	& 6\%	& 94\% \\
\hline
Listen to audio of the author	        & 1\%	& 9\%	& 90\% \\
\hline
Automated contextual help	            & 3\%	& 22\%	& 75\% \\
\hline
Search keywords on the tutorial timeline	& 0\%	& 20\%	& 80\% \\
\hline
Able to see the face of the author  	& 40\%	& 33\%	& 27\% \\
\hline
\end{tabular}
\end{table}
\subsection{PERFORMANCE TEST AND RESULTS}
We conduct a performance test using JMeter to measure the total system response time for a student to log in and start playback a programming tutorial. We aim to achieve at most 5 seconds total response time \cite{nielsen2001homepage} to retain users' attention using our system. We simulate 50 concurrent users per section and up to 10 sections of programming classes to address SQ3. Our microservices are executed on two AWS EC2 T2.micro instances with JMeter on a separate instance. All instances run in the same AWS virtual private cloud to minimize network latency variations in our tests.

Our first test uses 2 instances to host our microservices and manage to support up to 200 concurrent users within 5 seconds. Our second test increases the instances to four to support up to 500 users. As expected, the results show that 4 instances performed better than 2 for up to 200 concurrent users. For 300, 400 and 500 users, the results are still within our desired range of 5 seconds. We decide on 2 instances for up to 200 users and 4 instances for up to 500 users to balance between performance and hosting costs. Instance costs vary between 245.88 USD (2 instances) and 491.64 USD (4 instances) yearly with AWS auto-scaling. Costs are calculated based on pricing calculator \cite{AWSPricing}.

\section{CONCLUSION AND FUTURE WORK}
This paper describes our work to develop an extensible and scalable environment for integrated authoring and playback of interactive programming tutorials. This environment allows the authors to record bite-sized tutorial sections efficiently without custom software installation. We also propose a recording approach to record interactivity actions in text and audio artefacts instead of video frames to provide a more realistic learning experience and use these artefacts to contextual assist the student better. The results of our two studies show participants agree with the ease of recording when using our environment, feel that the increased interactivity can help learn to program better, and our environment has a good usability level based on System Usability Scale (SUS). We also conduct a performance test to show that our environment can support up to 500 concurrent users with a total system response time of less than 5 seconds.

We acknowledge the need to continue evaluating this environment and incorporate this environment usage into more courses. Besides working on the participant's suggestions, we also want to explore features in the area of learner analytics to better understand the students' behaviour for each tutorial.

\begin{acks}
This work is supported by Singapore Management University Educational Research Fellowship and Research Lab for Intelligent Software Engineering. We also like to express our gratitude to the six undergraduates (Emmanuel Tan Sheng Wei, Frans Corfiyanto, Claire Liew Xi Wei, Lee Yi Hao, Sim Sheng Qin, Brennan TAN Jinle) who helped to develop this environment and the participants of our two user studies.
\end{acks}

\bibliographystyle{ACM-Reference-Format}
\bibliography{ITSS-base}










\end{document}